# Native American Flute Ergonomics

## Clinton F. Goss, Ph.D.

Westport, CT, USA. Email: clint@goss.com

## ABSTRACT

This study surveyed ergonomic issues in 308 Native American flute players. It also correlated the physical measurements of a subgroup of 33 participants with the largest flute they found comfortable. The data was used to derive a predictive formula for the largest comfortable flute based on physical measurements. The median age of players was 63 years with a mean of 6.9 years playing Native American flute. Females reported significantly less time playing the instrument ($p$ = .004), but significantly faster self-reported progress rates ($p$ = .001). Physical discomfort was experienced by 47–64% of players at least some of the time. Over 10% of players reported moderate discomfort on an average basis. Females report significantly higher maximum and average physical discomfort than males ($p <$ .001 and $p =$ .015, respectively). Height, arm span, hand span, and reported length of time playing and experience level all correlated with the largest flute that the player found comfortable. Multivariate coefficient analysis on those factors yielded a formula with a strong correlation to the largest comfortable flute ($r =$ +.650). However, the formula does not have sufficient correlation to have value in predicting flute design. Customization of Native American flutes with the goal of improving ergonomics is proposed as a worthwhile goal.

## Introduction

The ergonomic issues particular to playing musical instruments have been widely studied. However, most research studies have focused on orchestral instruments and trained, experienced musicians (for example, [Lonsdale 2014] for Western concert flute, [Furuya 2006] and [Boyle-RB 2010] for piano, and [Wade-MA 2008] for trumpet). The bulk of the assistance and advice published for musicians and music teachers is similarly focused on experienced musicians and orchestral instruments (for example, [Norris 1993], [Horvath 2010], [Guptill 2010], and [Musikerhalsan 2014]).

This exploratory pilot study focuses on the ergonomics of playing the Native American flute, an ethnic wind instrument with roots in indigenous North American cultures. In contrast with orchestral instruments, the designs of Native American flutes are highly variable, with a very wide range of pitches, tunings, fingerings, temperament, and playing characteristics. The community of players of the instrument has relatively less training and experience than orchestral instrument players, and the focus of music education tends to be on improvisation and self-expression rather than precise renditions of written music. These differences have contributed to the recent popularity of the Native American flute in community music settings.

Because of the relative freedom in design and construction of the instrument, there is an opportunity to shift the focus for ergonomically comfortable instruments from the player (eg. exercises to increase reach and hand flexibility) to the maker of the instrument, even to the point of allowing individually customized instruments for a given player.

The first goal of this study was to survey the characteristics of players of Native American flutes, especially those related to ergonomic issues, the prevalence of discomfort, and the relationships to specific medical conditions. The second goal was to investigate the feasibility of using a system based on body measurements (in particular, a systems amenable to self-measurement by the player) to predict the limits of flute geometry that a given player would find comfortable.

This paper also provides incidental statistics of a more general nature regarding the characteristics of the Native American flute players who participated in this study.

**The Instrument**

The Native American flute is a "*front-held, open-holed whistle, with an external block and internal wall that separates a mouth chamber from a resonating chamber*" (R. Carlos Nakai, personal communication, June 21, 2002, as cited in [Goss 2011]). The instrument first appeared in the historical record in the early 19th century, and has been known by various names such as "courting flute", "love flute", "plains flute", "woodlands flute", and "śi'yotaŋka" ([Densmore 1918]).

The Native American flute is classified in the same family as the recorder.[1] It uses a duct or flue to direct the player's airstream, allowing the instrument to be played without the need for players to learn to form an embouchure with their lips. It is distinguished from the recorder by the inclusion of a slow air chamber which precedes the flue, providing an air reservoir that acts as a modest pressure bladder, tending to smooth out changes in breath pressure. Another distinguishing characteristic is its limited pitch range – typically no more than 1.3 octaves from the lowest note on the instrument.

Compared with other woodwind instruments, the sound chamber of most Native American flutes has a larger diameter relative to its length. This allows the instrument to maintain a full sound through a relatively large range of breath pressures without jumping registers (Brent Haines, personal communication, September 12, 2014).

Figure 1 shows the typical elements used in the design of a Native American flute. Since there are no common design standards, contemporary instrument makers take far more freedom in their designs than makers of orchestral wind instruments.

Figure 2 highlights some of the ergonomic issues related to the Native American flute. Since the instrument uses open holes, there are no key mechanisms to extend the reach of the fingers. Lower-pitched instruments require correspondingly long sound chambers, resulting in finger holes further from the head end of the instrument and extending the required reach. Flute makers tend to use a relatively short slow air chamber to compensate for the longer sound chamber.

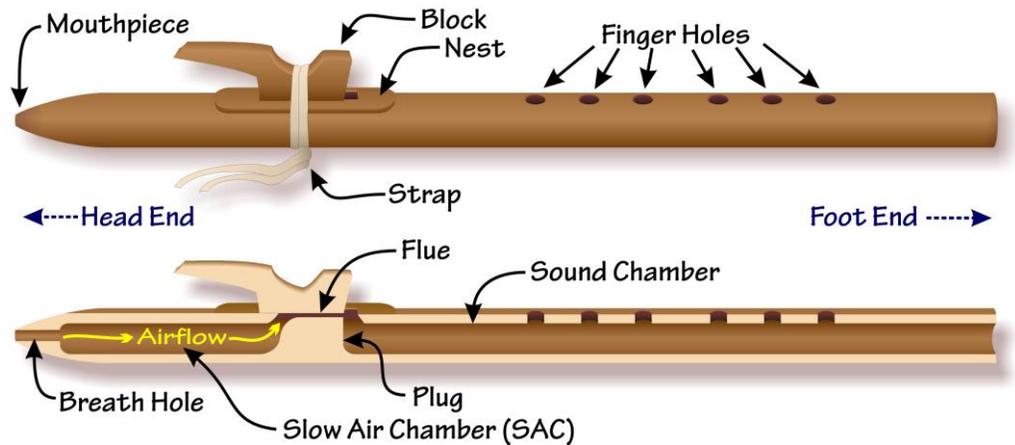

**Figure 1**. Anatomy of the Native American flute.

For those lower-pitched instruments, the best location of finger holes from an acoustic perspective would cause prohibitively large spread in the fingers of each hand, so compromises between ergonomics and acoustics are often made by the flute maker.

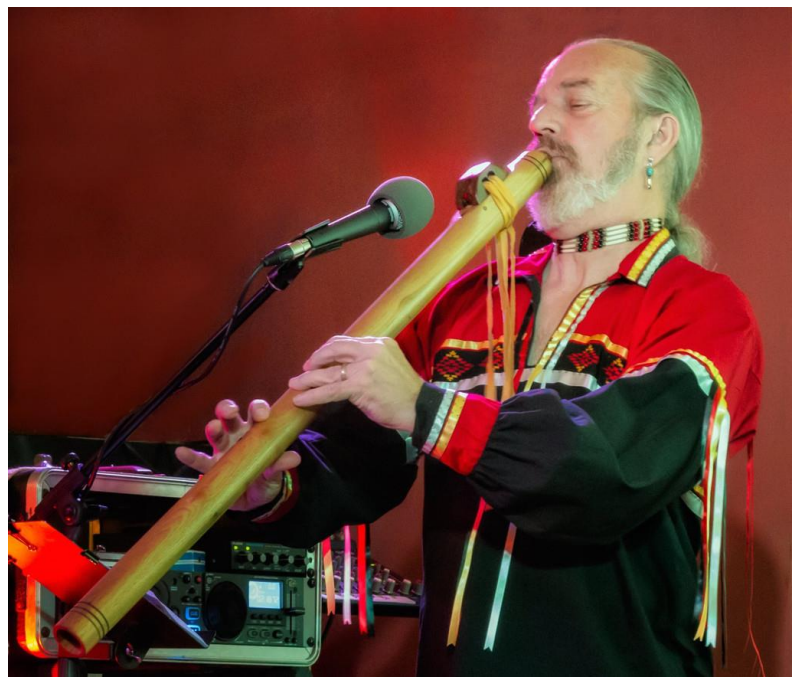

**Figure 2**. Ergonomic issues related to playing a low-pitched Native American flute. Photo courtesy of Randy "Windtalker" Motz.

---

[1] In the widely-used classification system of Hornbostel & Sachs (1914).

## Methods

A convenience sample of 308 participants was recruited from two sources, comprising two subgroups in the subsequent analysis:

- **On-line subgroup:** ($n = 275$) Members of on-line social network groups related to playing Native American flutes were recruited to complete an on-line questionnaire.

- **In-person subgroup:** ($n = 33$) Participants of a weeklong Native American flute workshop were recruited to:

  - complete a paper questionnaire,

  - get photographed for measurement purposes, and

  - play flutes of various sizes to locate flutes that were the "largest comfortable" and "slightly larger than comfortable" (i.e. "slightly uncomfortable") and complete a questionnaire on the specifics of comfort issues playing those flutes.

All participants signed or otherwise affirmed via an on-line form participation in the study and receipt of an informed consent. No coaching or recommendations regarding ergonomics were provided to participants prior to completion of their participation in the study.

### Questionnaire

The information gathered in the paper questionnaire and the on-line questionnaire was substantially the same. Completed questionnaires were visually inspected to remove accidental and duplicate submissions. One participant indicating transitional gender was retained, but eliminated from all analysis dealing with gender.

The information gathered from the questionnaire is shown in Figure 6 in the appendix. "Physical discomfort" was rated on a modified version of the Numeric Pain Rating Scale (NPRS) described in [McCaffery 1993] and shown in Figure 3.

### Body Measurement

Overall body height was gathered from all participants by self-report on the questionnaire.

Other physical measurements were taken of participants in the in-person subgroup by photographing the participants juxtaposed against a measuring scale and determining various physical dimensions from those photographs at a later time. These additional measurements comprised:

- **Arm span** between the tips of the longest fingers, measured against a wall, with outstretched arms raised parallel to the ground at shoulder height;

- Left and right **forearm length**, measured from the tip of the longest finger to a solid block against which the elbow and upper arm were placed; and

- Hand measurements, taken from photographs of each hand pressed firmly against a printed scale developed for this study (Figures 7 and 8 in the appendix).

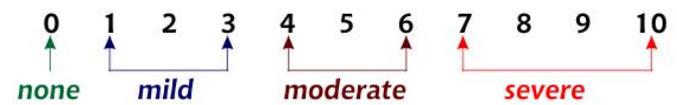

**Figure 3.** Modified Numeric Pain Rating Scale (NPRS).

The system of measurement based on photographs was used so that additional measurements could be obtained retrospectively as needed. All hand measurements were obtained by measuring the physical distance between two points on the photograph as it was displayed on a flat-screen LCD display with a 1:1 pixel aspect ratio (Figure 4). The physical distance measured on the display was scaled by a corresponding measurement of the physical distance on the display of the underlying scale on the paper.

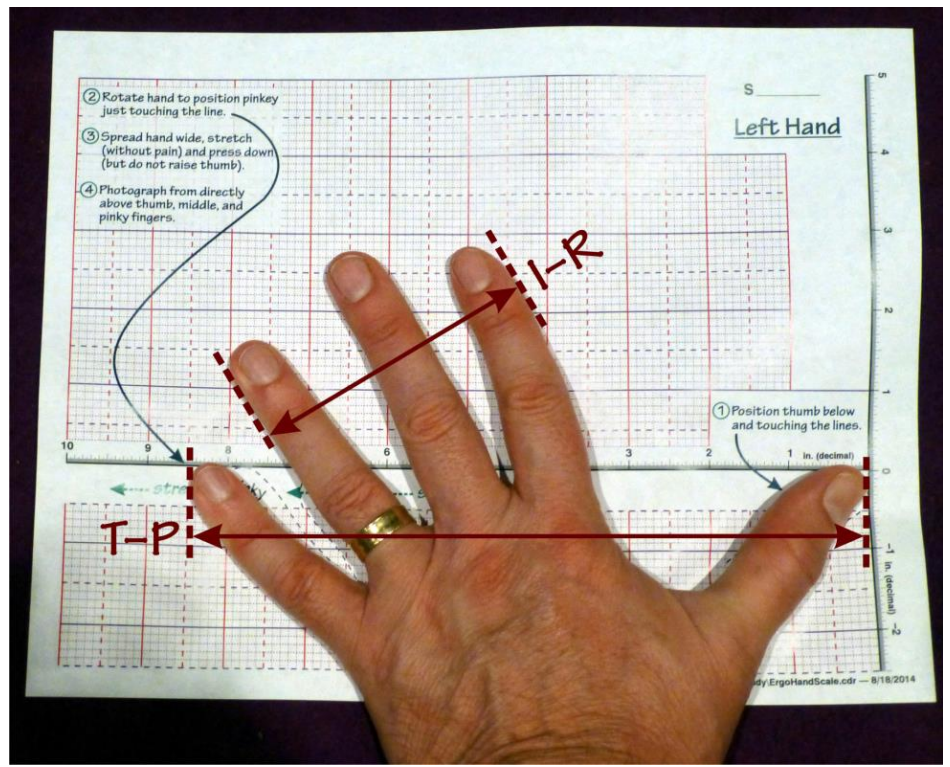

**Figure 4.** Hand Measurement.

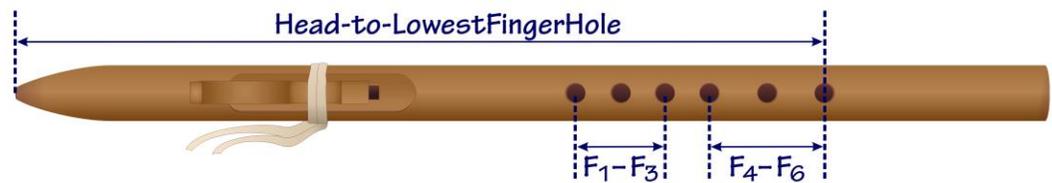

**Figure 5.** Flute measurement parameters.

The specific hand measurements in this study are:

- **T–P**: Thumb to pinky, measured from the outside edges of those two fingers.

- **I–R**: Index finger to ring finger, measured between the outside edges of those two fingers.

### Flute Measurements

This study provided an array of various sized flutes for the in-person group to play, in order to find the "largest comfortable flute". Three measurements were taken on each of those flutes (see Figure 5):

- **Head-to-LowestFingerHole:** The distance from the extreme head end of the flute to the center of the finger hole closest to the foot end of the flute.

- **$F_1$–$F_3$:** The maximum distance between the centers of the finger holes typically used by the hand closest to the head end of the flute.

- **$F_4$–$F_6$:** The maximum distance between the centers of the finger holes typically used by the hand closest to the foot end of the flute.

The metric **MaxIntraHandSpread** – the maximum distance that the fingers on either hand were required to span – was set to: max($F_1$–$F_3$, $F_4$–$F_6$).

For measurements that terminate at the center of a finger hole, two measurements were averaged: a measurement to the furthest edge of the finger hole and a measurement to the nearest edge of the finger hole.

### Largest Comfortable Flute

Participants in the in-person subgroup worked with the range of flutes that had been measured. Some participants found even the largest flutes to be well within their comfort range. However, 24 participants were able to locate flutes that were right at their limit of comfort, and completed a survey regarding that flute (Figure 9 in the appendix).

### Results

Data analysis was performed in Microsoft Excel 2010. All Student's t-tests assume two-tailed distribution and two heteroscedastic samples of unequal variance. Data analysis examined a wide range of measures and possible outcomes. Because multiple statistical inferences were considered simultaneously, the statistical measures presented in this study should be considered exploratory.

### On-line vs. In-person subgroups

To identify potential differences between questionnaire responses from the on-line subgroup and questionnaire responses from the in-person subgroup, pairwise t-tests across all questionnaire responses were performed. To eliminate inherent gender differences, separate male and female pairwise t-tests were performed for reported height, arthritis, and osteoporosis. These pairwise t-tests did not identify any significant differences between the on-line and in-person subgroups ($p > .200$ in all cases).

### Age and Height

Reported age ranged from 17 to 87 years with a mean of $61.97 \pm 10.03$ years and a median of 63 years. 300 of the 308 participants reported an age of 38 years or older. Males reported significantly higher age than females: 62.81 *vs.* 60.44 ($p = .044$).

Mean reported height was $64.14'' \pm 2.80''$ for females and $70.21'' \pm 2.92''$ for males.

**Table 1.** Participants by Reported Experience Level

| Reported Experience Level | Count | % of Total | Playing Time (years) |
|---|---|---|---|
| Beginner | 28 | 9.2% | 1.48 |
| Novice | 72 | 23.5% | 4.18 |
| Intermediate | 142 | 46.4% | 7.82 |
| Advanced | 62 | 20.3% | 10.55 |
| All | 306 | | 6.92 |

*Note*: Playing Time is the average of the reported time playing Native American flute by members of that experience level.

### Time Playing and Experience

Reported time playing all wind instruments ranged from 0 to 65 years with a mean of 14.30 years.

Reported time playing Native American flute ranged from 0 to 42 years with a mean of 6.92 years. 182 participants (59.2%) reported the same time for all wind instruments and the Native American flute, implying that the Native American flute was their first or only wind instrument.

The breakdown of participants by reported experience level – Beginner, Novice, Intermediate, or Advanced – is shown below in Table 1. Reported experience level has only a moderate positive correlation with reported time playing Native American flutes ($r = +.460$) and a weak positive correlation with reported time playing all wind instruments ($r = +.360$).

Males reported significantly more time playing Native American flute than females: 7.57 vs. 5.72 years, respectively ($p = .004$). However, there was no significant gender-based difference in reported experience level.

To analyze this relationship further, a composite metric called "progress rate" was developed. Progress rate is a numeric version of experience level (Beginner = 0, Novice = 100, Intermediate = 200, Advanced = 300) divided by the number of months playing Native American flute.

Females reported a significantly higher progress rate than males: 4.61 vs. 2.91, respectively ($p = .001$).

### Health Conditions

Some level of arthritis was reported by 176 participants (57.5%), with 120 reporting "mild", 45 reporting "moderate", and 11 reporting "severe" arthritis.

Osteoporosis was reported by 45 participants (14.9%), of which 33 were female, resulting in a very significant correlation between gender and osteoporosis ($p < .001$). The degree of osteoporosis was reported as "mild" by 34 participants, "moderate" by 10 participants, and "severe" by 2 participants (who were both male).

Gout was reported by 21 participants (6.8%), with 15 reporting "mild" and 6 reporting "moderate" gout.

Reported maximum physical discomfort ranged as high as 10 on the modified NPRS scale and averaged 2.30 across all participants. Average physical discomfort ranged up to 8 with an average of 1.06. Some physical discomfort was reported on an average basis by 144 participants (46.7%), with 27 participants reporting average physical discomfort in the moderate range and 3 in the severe range.

Physical discomfort was reported at least some percentage of the time by 198 participants (64.3%), with 32 participants reporting physical discomfort at least 50% of the time and 5 reporting physical discomfort 100% of the time.

Reports of maximum and average levels of physical discomfort, and percentage of time experiencing physical discomfort all showed a moderate positive correlation with the degree of arthritis ($r = +.423$ for average level of physical discomfort).

Females reported significantly higher maximum physical discomfort on the NPRS scale (3.03 vs. 1.90, $p < .001$) and average physical discomfort (1.36 vs. 0.89, $p = .015$).

Beginners and novices, as a subgroup, reported higher average physical discomfort than the subgroup of intermediate and advanced players (1.29 vs. 0.91), a result that approached significance ($p = .053$).

### Physical Measurements

Based on measurements of the in-person subgroup, measured arm span showed a very strong positive correlation with reported height ($r = +.938$). The average ratio of measured arm span to reported height was 98.34%.[2]

The correlation between reported height and measured left and right forearm length were somewhat less strong ($r = +.854$ and $r = +.870$, respectively). Left and right T–P measurement correlations with reported height were weaker ($r = +.633$ and $r = +.634$, respectively), as were left and right I–R measurements ($r = +.467$ and $r = +.465$, respectively).

The in-person subgroup was divided into "longer-time players" (participants with 5 years or more time playing Native American flutes – the median for the in-person subgroup) and "shorter-time players" (less than 5 years of time playing Native American flutes).

Longer-time players had significantly higher T–P (thumb–pinky) measurements than shorter-time players ($p = .007$ for the left hand and $p = .009$ for the right hand). That relationship persisted when the T–P were scaled by reported height ($p = .020$ for the left hand and $p = .026$ for the right hand).

The longer/shorter-time distinction was not a significant predictor of I–R (index–ring fingers) measurement of the left hand, either in absolute terms or as a percentage of reported height ($p = .302$ and $p = .626$, respectively). However, longer-time players did have significantly higher I–R absolute measurements for the right hand ($p = .049$, and $p = .100$ for I–R measurements scaled to reported height).

---

[2] These results confirm the observations by the ancient Roman architect Vitruvius and portrayed by Leonardo Da Vinci in the celebrated Vitruvian Man drawing and text.

### Largest Comfortable Flute

The metrics of Head-to-LowestFingerHole (H-LFH) and MaxIntraHandSpread (IHSmax) of the largest comfortable flute was moderately positively correlated with reported height ($r = +.470$ and $r = +.413$, respectively). H-LFH and IHSmax also showed a moderate positive correlation to time playing Native American flutes ($r = +.474$ and $r = +.509$, respectively) as well as reported experience level ($r = +.539$ for H-LFL and $r = +.593$ for IHSmax).

H-LFH and IHSmax showed various moderate positive correlations with measured arm span, forearm lengths, and various hand span measurements, but none were notably stronger than the correlation of H-LFH and IHSmax with reported height.

Arthritis showed virtually no correlation with H-LFH ($r = +.005$) and IHSmax ($r = +.006$). Osteoporosis showed similarly very weak correlations ($r = –.046$ for H-LFH and $r = +.021$ for IHSmax) as did gout ($r = +.169$ for H-LFH and $r = +.086$ for IHSmax).

### Coefficient Analysis

Multivariate coefficient analysis was used to explore the possibility of deriving a single formula that could reliably predict the H-LFH and IHSmax based on the parameters quantified by this study. Three coefficients were applied to the three parameters with the highest correlation to H-LFH and IHSmax: reported height, time playing Native American flutes, and reported experience level.

The tuple of three coefficients that maximized the correlation with H-LFH and (independently) IHSmax were then determined. The analysis yielded strong correlations of $r = +.650$ for H-LFH and $r = +.662$ for IHSmax.

The results of this formula were then compared with the H-LFH and IHSmax reported by the participants of this study. Although the formula shows strong correlations to with H-LFH and IHSmax, the formula produced errors as high as 5.95″ for H-LFH, demonstrating that it is not useful as a predictive tool.

### Discussion

### Age and Height

The study participants reported themselves to be substantially older than the general American population: median age 63 years vs. 37.5 years [Census 2013], with a mean reported age of $61.97 \pm 10.03$ years and the predominance of participants (300 of 308) who reported themselves to be older than the median American age. While no authoritative references as to the age of musicians could be located, it appears likely that the age profile of Native American flute players is substantially different than the age profile for players of orchestral instruments.

The mean reported height was taller than the average measured height for U.S. adults over 19 years of age as reported in [McDowell 2008] for both females (64.14″ vs. 63.86″) and males (70.21″ vs. 69.41″). One explanation is the that self-reports of height are significantly greater than measured heights – [Danubio 2008] reported a difference of 2.8 cm (1.10″) for females and 2.1 cm (0.83″) for males, which exceed the differences found in this study in both males and females.

### Physical Discomfort

Participants reported average physical discomfort at the low end of the mild range with maximum physical discomfort averaging above the midpoint of the mild range and some participants reporting average physical discomfort in the severe range. By various metrics, 47–64% of players experience physical discomfort at least some of the time, with over 10% reporting at least moderate discomfort on an average basis.

Putting these reports of physical discomfort in context is difficult, given the lack of published data for comparable situations. [Culf 1998] (according to [Mitchell-T 2007]) reported that 64–76% of symphony orchestra musicians surveyed experienced repetitive strain injuries that affected their performance, but this provides only a very weak comparison due to dramatically different populations and conditions.

The significant bias toward higher physical discomfort reported by females vs. males may also be significant in light of the significant less time reported by females vs. males for playing Native American flutes. One explanation is that the higher level of physical discomfort increases the "drop-out rate" playing the instrument.

### Physical Measurements

While physical measurements of arm span correlated closely with reported height, the physical measurements of hand span from the thumb to the pinky for both hands showed a strong relationship to their time playing Native American flutes. Taken by itself, this relationship could indicate that time playing Native American flute causes a wider hand span, or it could indicate that players with a wider hand span (either in absolute terms or in relationship to their height) are more apt to continue playing the instrument. However, when we consider that:

- the corresponding relationship between the index and ring fingers (significantly greater spread for in-person

participants with five or more years of playing Native American flutes) holds for only the right hand and not the left hand, and that

- the right hand is typically the hand used on the finger holes closer to the foot of the flute and that those holes tend to be further apart, it follows that playing Native American flute increases the ability to stretch the hand.

The methods used in this study to measure hand span were chosen so that a flute player could photograph their own hands against a common grid and flute makers could then reliably evaluate hand span from those photographs. However, while this approach produced useful comparative measurement within the in-person subgroup of participants, those hand measurements may not relate to the hand span used by the player on the cylindrical body of the flute.

### Primary Questions Investigated in this Study

In light of the results of this study, no clear relationship was found between a physical metric and the layout of finger holes on the largest comfortable flute. It may be that the limits of comfortable finger hole layout correlate with physical parameters not quantified by this study, or that there are other non-quantitative attributes of flute players that affect comfort.

This result was discussed with several experienced makers of Native American flutes, who concurred with this finding based on their own experience.[3]

### Limitations

Some general limitations in the design and execution of this study include:

- The small number of participants in the in-person subgroup ($n = 33$) may have limited the significance of some the results and caused type II errors in statistical testing of the hypotheses of this study.

- The use of a convenience sample rather than statistical sampling may have skewed some results. The participants may not have been a representative sample of the population of Native American flute players.

- Only flutes with finger holes positioned along the centerline of the instrument were used. Some Native American flutes, especially lower-pitched instruments, have finger holes offset from the centerline to improve comfort and reachability.

- Participants in the in-person subgroup had a relatively short time playing the range of flutes available to determine the largest comfortable flute. It may be that the discomfort experienced by a player on a given flute changes over days or weeks as experience is gained playing that flute.

A number of difficulties were noted during the process of measuring the hands, and can provide insight to future studies using a similar strategy for measuring physical characteristics:

- The direction given to participants to "fully spread your hand" resulted in various degrees of effort in stretching. The effort was not necessarily associated with the amount of stretch used while playing.

- Photographic parallax is an issue, since the measurement point was above the surface of the graph paper – closer to the camera. This would tend to decrease the measurement of the underlying scale on the graph paper relative to the quantity measured and make all measurements higher than the actual distances. This issue was further compounded by the lack of a standard height for the camera. Future studies could establish a standard height and make it at a substantial distance from the hand and graph paper to reduce the impact of photographic parallax.

- The presence of fingernails of various lengths created some ambiguity in the correct endpoint for measurements.

- The presence of shadows made it difficult to precisely locate the measurement endpoints in a few cases.

- Some participants rotated various fingers during the measurement process, creating a lack of uniformity across participants in the measurement process.

- In one case, the hand and paper were placed on a soft surface rather than on a hard table top, creating a slight undulation in the paper as the hand was pressed down.

Other limitations include:

- The use of self-reported rather than measured height may have affected the results.

- The self-reporting of experience level is particularly suspect. Some relatively objective evaluation of playing level may be preferable.

- The different formats for the questionnaires completed by the on-line subgroup and the in-person subgroup (on-line Web form *vs.* paper form) could have affected results, even though none were detected by the set of pairwise t-tests.

---

[3] One experienced flute maker, Brent Haines, commented that his experience is that the size of flute that can be comfortably accommodated is related more to the enthusiasm of the player than to physical measurements.

## Conclusions

This study was motivated by a lack of research in the area of ergonomic issues in the community of Native American flute players and the need for a straightforward system based on physical measurements whereby flute makers could construct custom flutes that would be unlikely to cause physical discomfort.

The community of Native American players is dramatically different from that of orchestral wind instrument players. Players of orchestral instruments report of a range of repetitive strain injuries and playing-related musculoskeletal disorders, and a surprisingly high percentage of musicians report playing in physical discomfort on an ongoing basis. Orchestral instruments are relatively fixed in design and effort is often placed on small modifications to accommodate ergonomic issues without changing the acoustic or mechanical properties of the instrument (e.g. [Storm 2006]).

> "*Musical instruments are hardly designed to be 'friendly'; rather, they are designed to achieve the best fit with a highly skilled human physiology*"
>
> ([Bernardini 2010])

In contrast, the community of Native American flute players has:

- A relatively older population focused on music for personal enjoyment and self-actualization;
- A focus on community music making rather than performance;
- A culture of improvisation rather than playing written music;
- Instruments that are relatively free in design;
- A large population of Native American flute instrument makers, many of whom are ideally set up to craft custom-designed instruments that could maximize comfort.

In what seems like an ideal situation for minimizing physical discomfort, this study has found that many players still experience substantial physical discomfort. This may be due to the expectation (as with orchestral musicians) that physical discomfort is "normal" when playing any musical instrument.

This study also established a link between the level of physical discomfort and shorter time playing the instrument, and provided evidence that physical discomfort may cause an increase in the "drop-out" rate of players.

An informal survey of the marketing literature of flute makers shows many who emphasize the woods used, the precision of tuning, the tonal quality, and the spiritual aspects associated with the traditional roots of the instrument. Few mention the opportunity for an instrument (customized or otherwise) that addresses playing comfort.

With regard to the goal of finding a system to assist flute makers in creating such instruments based solely on physical measurements, this study did not produce definitive results. It may be that the process of designing an instrument that will be comfortable for a player requires in-person examination and consultation by a person experienced in flute design and/or issues of ergonomics.

It appears that a greater focus on the design of instruments that address ergonomic issues would benefit the community of Native American flute players.

## References

For a general bibliography in the areas covered by this article, see http://www.Flutopedia.com/refs_ergo.htm.

[Bernardini 2010] Nicola Bernardini. The Role of Physical Impedance Matching in Music Playing, Sound is Motion Symposium, Stockholm, Sweden, February 11, 2010, 2010.

[Boyle-RB 2010] Rhonda B. Boyle and Robin G. Boyle. "Hand Size and the Piano Keyboard. Technical and Musical Benefits for Pianists Using Reduced-Size Keyboards", Journal of the Victorian Music Teachers' Association (VMTA), Volume 36, Number 1, March 2010, pages 17–35.

[Census 2013] U.S. Census Bureau. *Comparative Demographic Estimates – 2013 American Community Survey 1-Year Estimates*, Publication CP05, http://factfinder.census.gov/, retrieved December 22, 2014.

[Culf 1998] Nicola Culf. *Musicians' Injuries: A Guide to Their Understanding and Prevention*, Parapress Ltd, Tunbridge Wells, U.K., ISBN 1-898594-62-7.

[Danubio 2008] Maria Enrica Danubio, Gaetano Miranda, Maria Giulia Vinciguerra, Elvira Vecchi, and Fabrizio Rufo . "Comparison of Self-reported and Measured Height and Weight: Implications for Obesity Research among Young Adults", Economics & Human Biology, Volume 6, Issue 1, March 2008, pages 181–190, doi:10.1016/j.ehb.2007.04.002.

[Densmore 1918] Frances Densmore. *Teton Sioux Music and Culture*, Smithsonian Institution, Bureau of American Ethnology, Bulletin 61, published by the United States Government Printing Office, Washington, D.C., 1918, 561 pages.

[Furuya 2006] Shinichi Furuya, Hidehiro Nakahara, Tomoko Aoki, and Hiroshi Kinoshita. "Prevalence and Causal Factors of Playing-Related Musculoskeletal Disorders of the Upper Extremity and Trunk among Japanese Pianists and Piano students", Medical Problems of Performing Artists, Volume 21, Number 3, 2006, pages 112–117.

[Goss 2011] Clinton F. Goss. *Anatomy of the Native American flute*. Flutopedia: http://www.Flutopedia.com/anatomy.htm. Retrieved April 12, 2012.

[Guptill 2010] Christine Guptill and Christine Zaza. "Injury Prevention: What Music Teachers Can Do", Music Educators Journal, Volume 96, Number 4, published by MENC: The National Association for Music Education, June 2010, pages 28–34.

[Horvath 2010] Janet Horvath. *Playing (Less) Hurt — An Injury Prevention Guide for Musicians*, Hal Leonard, 2010, 256 pages, ISBN-13 978-1-4234-8846-0, ASIN 1423488466.

[Lonsdale 2014] Karen Anne Lonsdale, E-Liisa Laakso, and V. Tomlinson. "Contributing Factors, Prevention, and Management of Playing-related Musculoskeletal Disorders among Flute Players Internationally", Medical Problems of Performing Artists, Volume 29, Number 3, September 2014, pages 155–162.

[McCaffery 1993] Margo McCaffery and Alexandra Beebe. *Pain: Clinical Manual for Nursing Practice*, published by the V. V. Mosby Company, Baltimore, Maryland, 1993.

[McDowell 2008] Margaret A. McDowell, Cheryl D. Fryar, Cynthia L. Ogden, and Katherine M. Flegal. "Anthropometric Reference Data for Children and Adults: United States, 2003–2006", National Health Statistics Reports, Number 10, October 22, 2008, 48 pages.

[Mitchell-T 2007] Tamara Mitchell; Sally Longyear (editor), *A Painful Melody: Repetitive Strain Injury among Musicians*, 2007, monograph retrieved December 17, 2014 from http://www.memphis.edu/music/pdf/painful.pdf.

[Musikerhalsan 2014] Artist-och Musikerhälsan. *Musician Ergonomics*, 2014, retrieved August 24, 2014 from http://www.artist-musikerhalsan.se/.

[Norris 1993] Richard N. Norris. *The Musician's Survival Manual — A Guide to Preventing and Treating Injuries In Instrumentalists*, First edition, published by the International Conference of Symphony and Opera Musicians (ICSOM), 1993, 134 pages, ISBN 0-918812-74-7 (978-0-918812-74-2).

[Storm 2006] "Assessing the Instrumentalist Interface: Modifications, Ergonomics and Maintenance of Play", Physical Medicine and Rehabilitation Clinics of North America, Volume 17, 2006, pages 893–903, doi:10.1016/j.pmr.2006.08.003

[Wade-MA 2008] Mark Alan Wade. *An Annotated Bibliography of Current Research in the Field of the Medical Problems of Trumpet Playing*, D.M.A. Dissertation – Ohio State University, 2008, xi + 121 pages.

## Acknowledgements

My gratitude goes to the participants in this study, and especially to the research assistants, Frank L. Henninger, Patricia B. Smith, and Tchin, who provided extensive support for the in-person subgroup. My thanks goes to Edward Kort, Brent Haines, and Judy Robinson for their helpful suggestions on this paper.

## Appendix – Forms

The following pages show the paper-based forms used for the in-person subgroup of this study.

## Ergonomics of Flute Playing — Questionaire                    S______

**Your Name:** ________________________________________________

**Today's date:** __________________  **Your age:** _________  **Your height:** ____ ft. ____ in.

**Your Sex (circle):    Female    Male    It's Complicated**

**Time playing any wind instrument:    ______ years ______ months**

**Experience playing Native flutes (circle):  Beginner  Novice  Intermediate  Advanced**

**Time playing Native American flutes:    ______ years ______ months**

**Relevant health issues (*particularly involving muscles, joints, bones, or nervous sytem*):**

| **Arthritis:** ☐ mild ☐ moderate ☐ severe | **Osteoporosis:** ☐ mild ☐ moderate ☐ severe | **Gout:** ☐ mild ☐ moderate ☐ severe |

☐ **Physical trauma (arms, hands, or neck — *please provide details*)**
☐ **Neuromuscular disease (*please provide details*)**

 **Other:** _________________________________________________

 **Details:** _________________________________________________

**<u>Maximum</u> level of physical discomfort you experience when playing Native flutes:**

        circle a number:   0   1   2   3   4   5   6   7   8   9   10
                            *none*   *mild*      *moderate*      *severe*

**<u>Average</u> level of physical discomfort you experience when playing Native flutes:**

        circle a number:   0   1   2   3   4   5   6   7   8   9   10
                            *none*   *mild*      *moderate*      *severe*

**What percentage of the time do you experience any level of physical discomfort when playing Native flutes: __________ %**

**Please describe any particular issues regarding discomfort playing flutes, or any other information you feel is relevant (use the back of this page, if needed):**

*Ergonomics of Flute Playing — C:\Act\ErgonomicsStudy\ErgoSurvey.cdr — 8/19/2014*

**Figure 6**. Survey form.

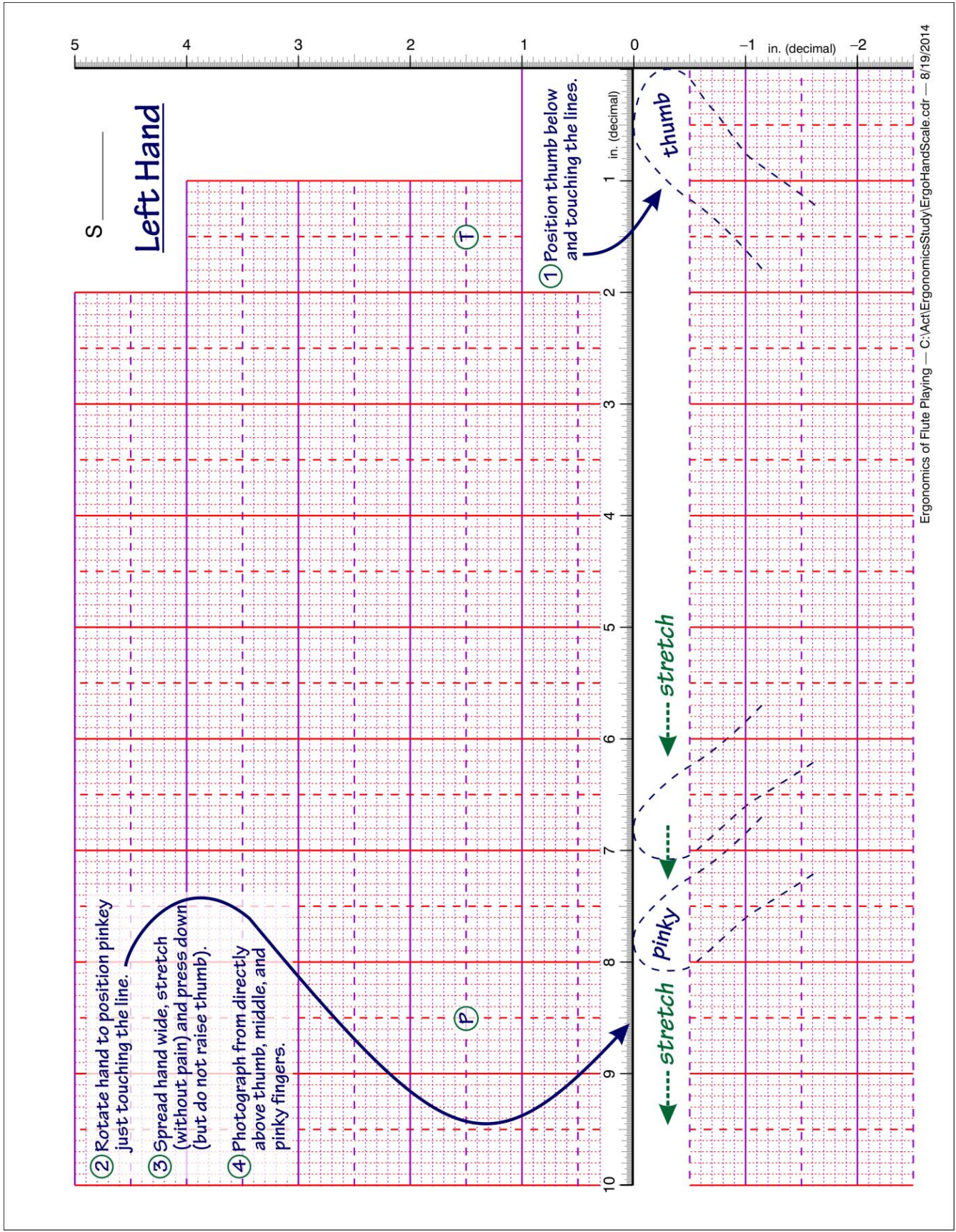

**Figure 7**. Form for left hand measurement.

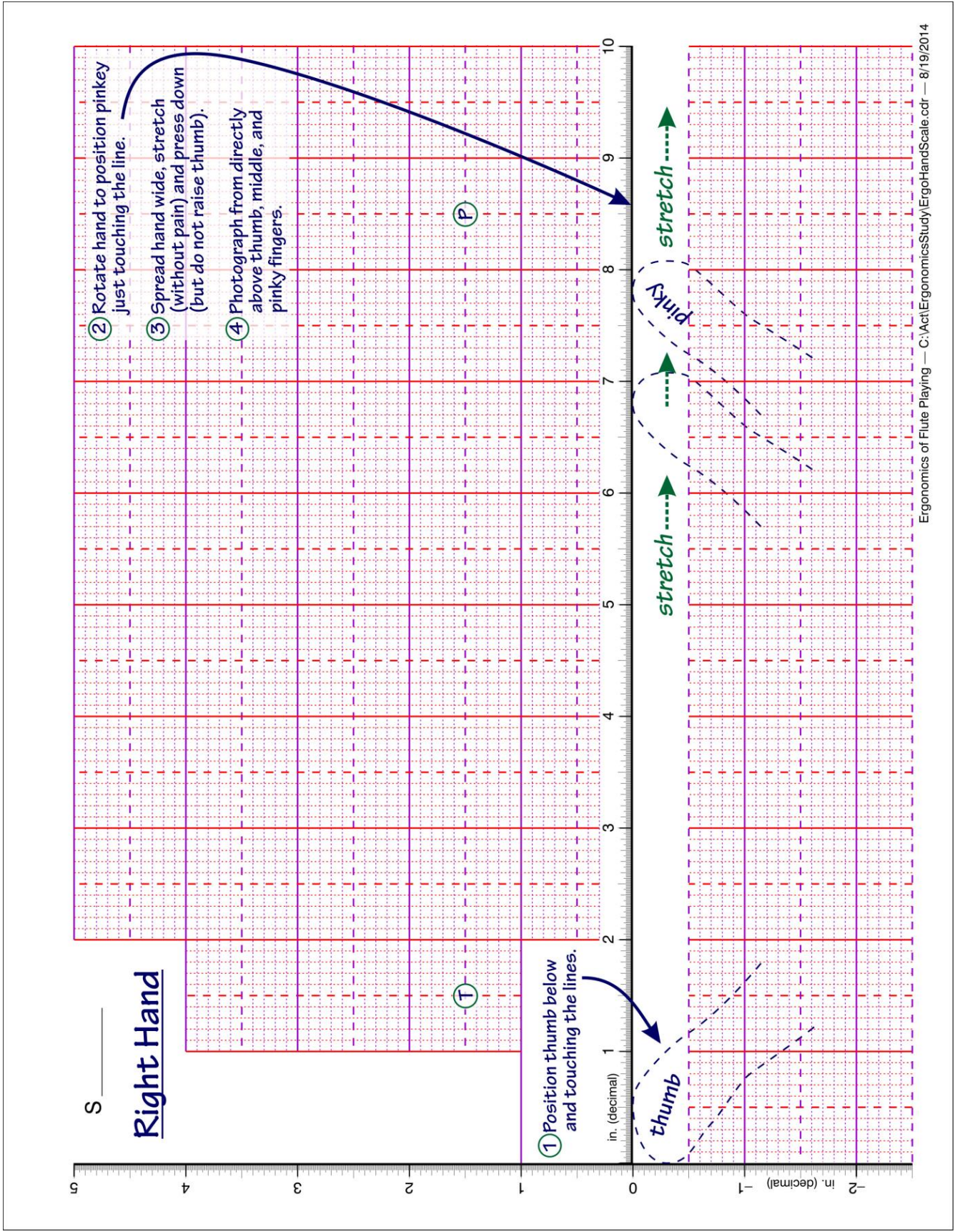

**Figure 8**. Form for right hand measurement.

<u>**Ergonomics Survey**</u>

S: __________     Flute ID: ____________     Today's Date: ________________________

Survey with this flute (circle):   First     Second     Third

Have you attended a workshop session on grip or finger reach? (circle):   Yes    No

Have you had private coaching on improving your finger reach? (circle):   Yes    No

Which hand is "on top" (covers finger holes nearest your mouth)? (circle):  Left  Right

<u>**Physical Discomfort Areas**</u>

| | none | mild | | moderate | | severe | |
|---|---|---|---|---|---|---|---|

| | none ↓ | mild |← | moderate ↓ | ↓ | severe ↓ | |
|---|---|---|---|---|---|---|---|
| Neck | 0 | 1 | 2 3 | 4 5 | 6 | 7 8 9 | 10 |
| Left shoulder | 0 | 1 | 2 3 | 4 5 | 6 | 7 8 9 | 10 |
| Left wrist | 0 | 1 | 2 3 | 4 5 | 6 | 7 8 9 | 10 |
| Left thumb | 0 | 1 | 2 3 | 4 5 | 6 | 7 8 9 | 10 |
| Left fingers | 0 | 1 | 2 3 | 4 5 | 6 | 7 8 9 | 10 |
| Right shoulder | 0 | 1 | 2 3 | 4 5 | 6 | 7 8 9 | 10 |
| Right wrist | 0 | 1 | 2 3 | 4 5 | 6 | 7 8 9 | 10 |
| Right thumb | 0 | 1 | 2 3 | 4 5 | 6 | 7 8 9 | 10 |
| Right fingers | 0 | 1 | 2 3 | 4 5 | 6 | 7 8 9 | 10 |

*none*        *mild*             *moderate*            *severe*

**Circle the finger holes that could not be completely covered (i.e. "leaked"):**

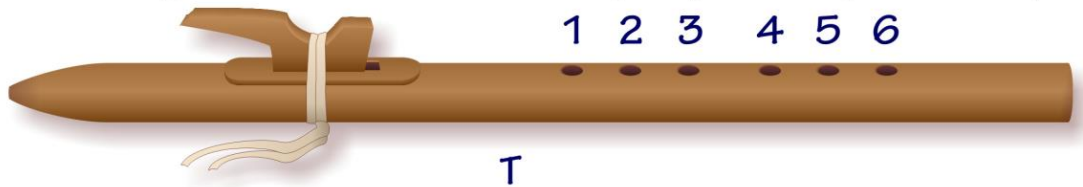

1  2  3    4  5  6

T

**Please provide any other thoughts / observations:**

**Figure 9**. Survey form.